\documentclass[conference]{IEEEtran}
\IEEEoverridecommandlockouts
\usepackage{cite}
\usepackage{amsmath,amssymb,amsfonts}
\usepackage{multicol}
\usepackage{graphicx}
\usepackage{textcomp}
\usepackage{xcolor}
\usepackage{enumitem}
\usepackage{caption}
\usepackage{subcaption}
\usepackage{multirow}
\usepackage{tabularx}
\usepackage{url}
\usepackage{algorithm,amsmath}
\usepackage{algpseudocode}

\makeatletter

\newcounter{phase}[algorithm]
\newlength{\phaserulewidth}
\newcommand{\setphaserulewidth}{\setlength{\phaserulewidth}}

\makeatother

\setphaserulewidth{.7pt}

\begin{document}

\begin{titlepage} 
 \textcopyright 2020 IEEE.  Personal use of this material is permitted.  Permission from IEEE must be obtained for all other uses, in any current or future media, including reprinting/republishing this material for advertising or promotional purposes, creating new collective works, for resale or redistribution to servers or lists, or reuse of any copyrighted component of this work in other works.
 \end{titlepage}

\title{Long-term Continuous Assessment of SRAM PUF and Source of Random Numbers}

\author{\IEEEauthorblockN{Rui Wang, Georgios Selimis, Roel Maes, Sven Goossens}
\IEEEauthorblockA{Intrinsic-ID, Eindhoven, The Netherlands} 
Email: \{rui.wang, georgios.selimis, roel.maes, sven.goossens\}@intrinsic-id.com}

\maketitle

\begin{abstract}
	The qualities of Physical Unclonable Functions (PUFs) suffer from several noticeable degradations due to silicon aging. In this paper, we investigate the long-term effects of silicon aging on PUFs derived from the start-up behavior of Static Random Access Memories (SRAM). Previous research on SRAM aging is based on transistor-level simulation or accelerated aging test at high temperature and voltage to observe aging effects within a short period of time. In contrast, we have run a long-term continuous power-up test on 16 Arduino Leonardo boards under nominal conditions for two years. In total, we collected around 175 million measurements for reliability, uniqueness and randomness evaluations. Analysis shows that the number of bits that flip with respect to the reference increased by 19.3\% while min-entropy of SRAM PUF noise improves by 19.3\% on average after two years of aging. The impact of aging on reliability is smaller under nominal conditions than was previously assessed by the accelerated aging test. The test we conduct in this work more closely resembles the conditions of a device in the field, and therefore we more accurately evaluate how silicon aging affects SRAM PUFs. 
\end{abstract}

\begin{IEEEkeywords}
SRAM PUF, long-term, aging, evaluation, reliability, uniqueness, randomness, entropy
\end{IEEEkeywords}

\section{Introduction}\label{introduction}
The number of  Internet-connected devices worldwide is expected to run in the tens of billions by 2030 \cite{IEEE}. One of the key components in the security of an Internet-connected device is a device-unique cryptographic identity that can be verified by the cloud infrastructure. Existing methods for securely storing such an identity in a large number of devices often rely on keys stored in one-time programmable memory. This method does not scale to billions of devices in the Internet of Things (IoT). An alternative solution for secure initiation of a cryptographic identity is based on  Physical Unclonable Functions (PUFs), which utilizes deep submicron manufacturing process variation during the fabrication of the device. PUFs technology based on Static Random Access Memory (SRAM) has been commercially deployed in products by Microsemi \cite{Microsemi} and NXP \cite{NXP}. SRAM PUF is applicable to either generate reliable secure keys (reliability and uniqueness requirement) or provide random entropy to the device (randomness requirement) \cite{Den}.

In commercial products, the lifetime of the device is a significant concern. Deep into the physical level of a silicon IC, some circuit's parameters slowly and gradually change over time, leading to the change of properties on the IC. This change is defined as \textit{silicon aging}. It has been presented that silicon aging degrades the reliability of SRAM PUF \cite{Roel2} while it improves noise entropy over time \cite{DRBG}.

One of the methods to analyze silicon aging on SRAM PUF is circuit simulation at transistor level. Another option is to measure a silicon device at high temperature and operating voltage which accelerates the aging effect. In this paper, for the first time we propose to run the silicon device in nominal condition for a sufficient amount of time to observe real-time, non-simulated and non-accelerated aging effects.
\subsection{Related Work}

The concept of SRAM PUF was firstly proposed by Guajardo et al. \cite{PUF}. Kumar et al. studied the degradation of SRAM cells for 70nm and 100nm CMOS technology by performing transistor level simulation \cite{NBTI}. Aging of other technologies including Partially Depleted Silicon On Insulator (PDSOI) \cite {PDSOI} and FinFET \cite{Said1} are also investigated based on simulation. On the other hand, Maes et al. used accelerated aging on the silicon to demonstrate SRAM PUF reliability degrades for 65nm CMOS technology \cite{Roel2, Roel1}. 

\subsection{Our Contribution}
This paper presents the first experimental aging test of SRAM PUF under nominal condition continuously running for two years. The nominal aging test reflects the practical aging effect on a device in the field more accurately, since we do not expect constant stress conditions during the device's lifetime. Long-term continuous aging assessment of the SRAM PUF with respect to reliability, uniqueness and randomness has been conducted. For the assessment we used SRAMs of low cost generic use Commercial Off-The-Shelf (COTS) devices.  

The evaluation indicates that even after years of continuous use, SRAM PUF remains of sufficient quality for reliable and secure key generation scheme and random number generation. The reliability property is worsened to some extent. The impact, however, is less pessimistic than the equivalent aging effect derived from accelerated aging.

\subsection{Paper Outline}
Section \ref{background} provides the background of SRAM PUF technology, in particular the silicon aging effect on SRAM PUF cells. Section \ref{measurement} describes the measurement setup of our long-term aging experiment. Section \ref{evaluation} presents the evaluation result and discussion on the aging data from the measurement, and the paper is finally concluded in Section \ref{conclusion}.
\section{Background}\label{background}
\subsection{SRAM PUF technology and applications}\label{SRAM PUF}
The basic 6T-SRAM cell structure consisting of two cross-coupled inverters and access transistors (not shown) is shown in Fig. \ref{fig:SRAM}. The initial state $Q$ after a new power-up is determined by uncontrollable variations during manufacturing. The preferred power-up state is persistent and stable after manufacturing. On the other hand, a small quantity of SRAM cells are unstable and show a random preference at every power-up. Taking both stable and unstable SRAM cells into consideration, previous research has shown that the power-up pattern of SRAM memory is perfectly reproducible (with the proper application of error-correction codes), unique \cite{Roel1} and also capable of providing randomness \cite{RNG}. Therefore, SRAM PUF can be applied in the following scenarios.
\subsubsection{Secure Key generation}
The most common application of PUFs is secure key generation and storage. The cryptographic key is derived based on SRAM PUF and stored via helper data scheme. In this application, reliability and security are important properties for SRAM PUFs. 

Reliability means the response of SRAM PUFs is reproduciable with limited amount of bit error rate in predefined environmental and system conditions. Error correction codes can be designed to correct up to 25\% of bit error rate without reproduction failure \cite{Bin}. 

On the other hand, security means two properties. First, SRAM PUF should have sufficient entropy to prevent significant information leakage on the generated key. It implies that bias present on SRAM PUFs should be within the boundary. Current debiasing schemes can deal with 25\% / 75\% bias \cite{Roel4}. Second, the response of an SRAM PUF is unpredictable even given all the responses of other SRAM PUFs. It implies that SRAM PUF should have good uniqueness.
\subsubsection{True Random Number Generation}
True Random Number Generator (TRNG) provides an unpredicted seed to cryptographic systems. Electrical noise influences the initial state of unstable SRAM cells. By utilizing these noises in the circuits, SRAM PUFs can also serve as TRNG. In this application, a sufficient amount of bits should flip over multiple power-ups in order that SRAM PUFs can provide sufficient randomness at next power-up test.

\subsection{Aging Effect on SRAM Cells}\label{NBTI}
\begin{figure}[htbp]
	\centerline{\includegraphics[width=0.27\textwidth]{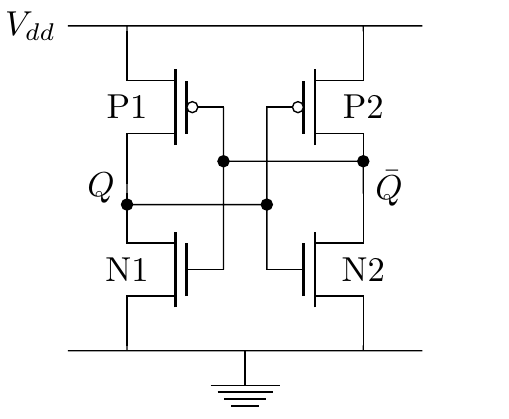}}
	\caption{Two cross-coupled inverters at the core of each SRAM cell (Other two access transistors not shown).}
	\label{fig:SRAM}
	\vspace{-3mm}
\end{figure}

During  nominal operations, a digital CMOS circuit in a silicon chip degrades over time. The dominant aging effect is Negative Bias Temperature Instability (NBTI) resulting in a temporal increase in the threshold voltage, particularly for a switched-on PMOS transistor \cite{Roel2}. For simplicity, all the following parameters related to PMOS are treated as positive values. In Fig. \ref{fig:SRAM}, when an SRAM cell stores state zero $(Q=0)$, P1 is switched off and P2 is switched on. As a result of NBTI, $V_{\mathrm{th,P2}}$ will increase while  $V_{\mathrm{th,P1}}$ is not affected. Also, the fact that P2 is switched on indicates that initially  $V_{\mathrm{th,P2}}<V_{\mathrm{th,P1}}$. Therefore, the temporal tendency of this SRAM cell is that $|V_{\mathrm{th,P2}}-V_{\mathrm{th,P1}}|$ grows smaller. This tendency also holds when $(Q=1)$.
 Smaller $|V_{\mathrm{th,P2}}-V_{\mathrm{th,P1}}|$ means the SRAM cell is more balanced and more likely to have PUF response bits flip. From the reliability perspective, this effect deteriorates SRAM PUF as key generation scheme since more SRAM cells tend to lose stability and randomly flip, i.e. the reliability of SRAM PUF worsens over time. From the randomness perspective, more SRAM cells become balanced and unpredictable over measurements, i.e. more randomness can be harvested based on SRAM PUFs.
 
Additionally, with the introduction of high-permittivity gate dielectrics, Positive Bias Temperature Instability (PBTI) for a switched-on NMOS transistor is also becoming more significant \cite{PDSOI}. Due to this effect, then NMOS transistor operating with a positive gate-to-source voltage experiences time-dependent threshold voltage increase. 

Other silicon aging mechanisms including hot electron injection, gate dielectric breakdown and electromigration are discussed in  \cite{Aging}.
\section{Measurement Setup}\label{measurement}
	The measurement platform was set up for the continuous SRAM PUF observation read-out at room temperature. SRAM chips were getting successive power cycles from Feb. 8, 2017 to Feb. 8, 2019. Each SRAM chip has been powered up and read out approximately 11 million times within this period, i.e. around 10 measurements per minute. For the more exhaustive evaluation, 16 devices were tested in the measurement setup, leading to around 175 million measurements in total.
	
	 The type of the SRAM chips used in the setup is the SRAM on an ATmega32u4 microcontroller on an Arduino Leonardo board. The operating voltage of ATmega32u4 is 5V, and the size of the SRAM is 2.5 KByte.
	\begin{figure}[ht]
		\centering
		\begin{subfigure}{0.33\textwidth}
			\centering
			\captionsetup{justification=centering,margin=0.1cm}
			\includegraphics[width=1.1\textwidth]{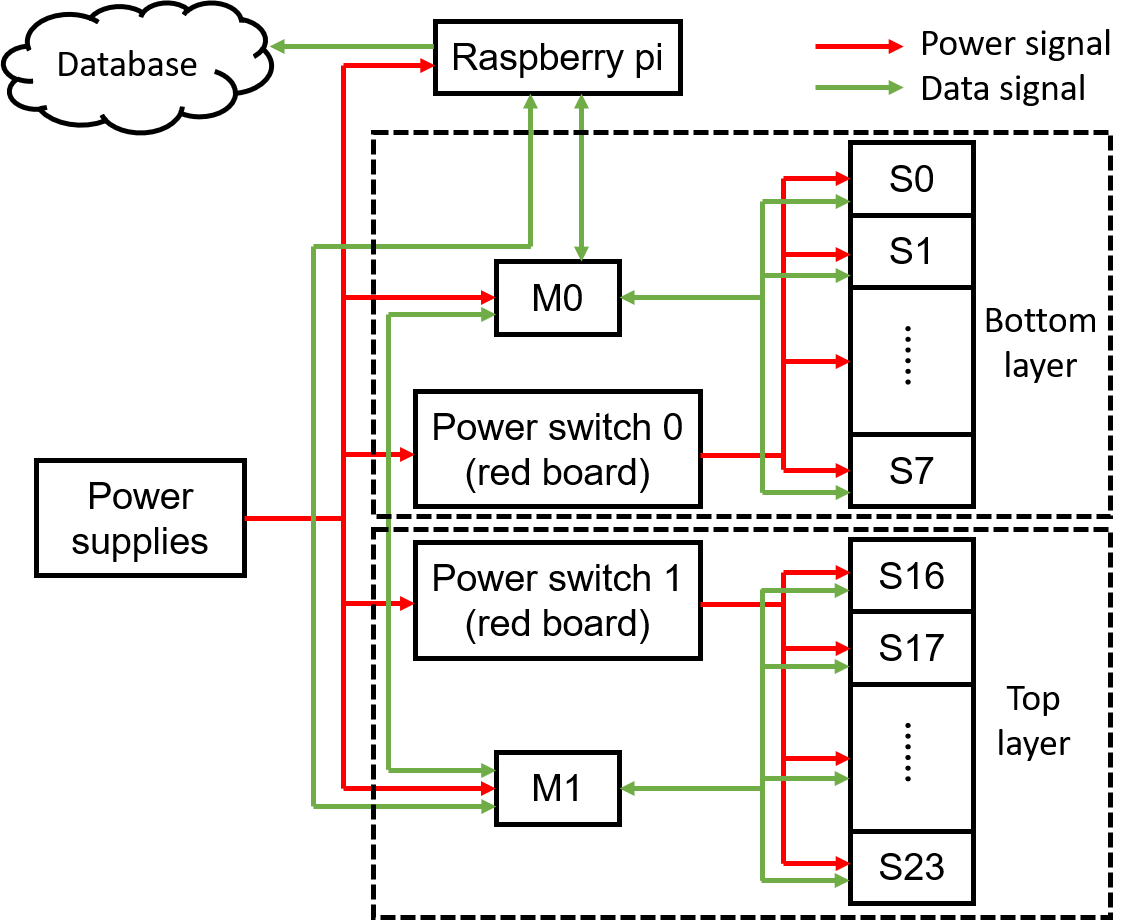}
			\caption{Schematic of SRAM PUF on Arduino boards measurement setup  and flow of power/data signals. }
			\label{fig:Measure1}
		\end{subfigure}	
		\begin{subfigure}{0.33\textwidth}
			\centering
			\captionsetup{justification=centering,margin=0.01cm}
			\includegraphics[width=1.1\textwidth]{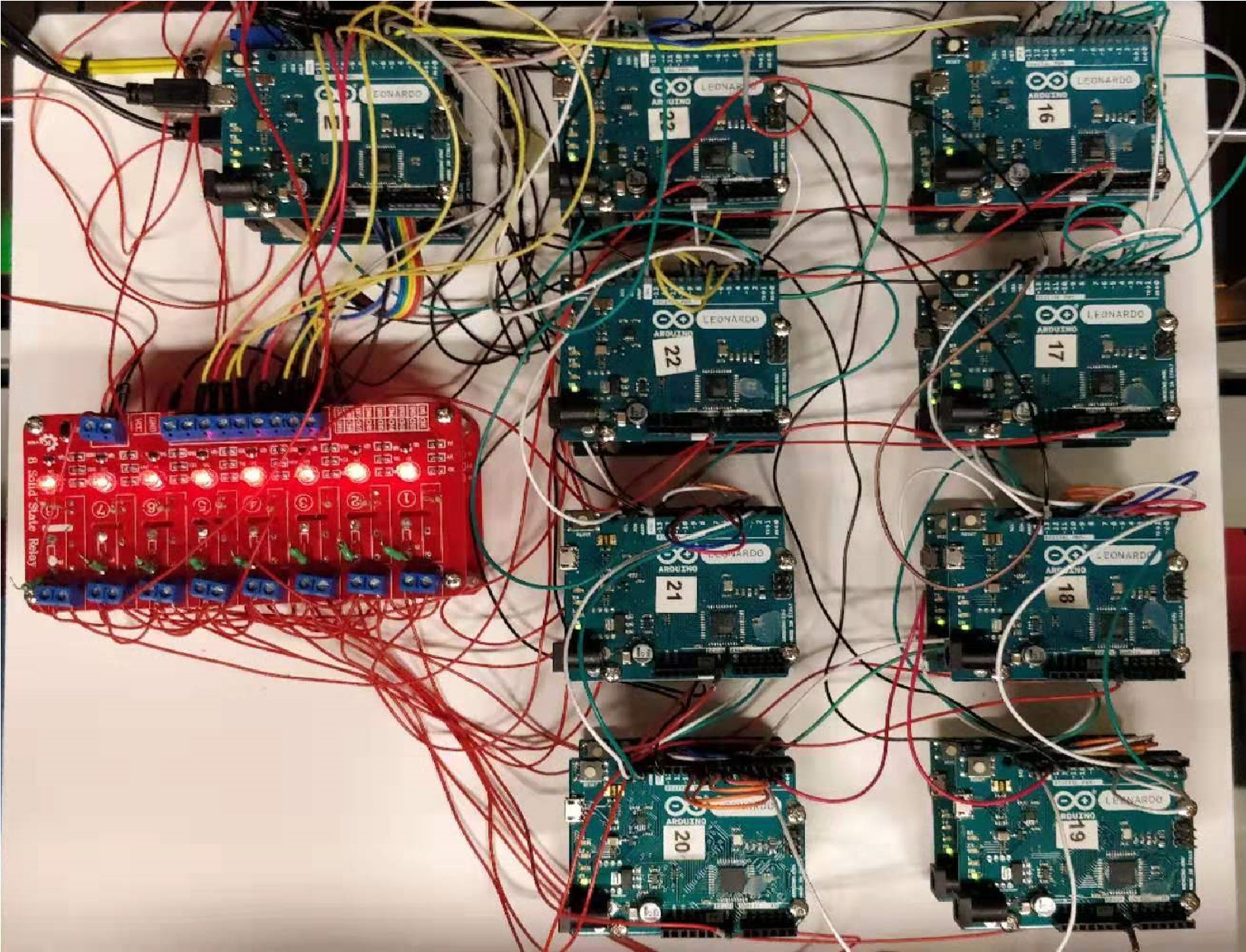}
			\caption{Practical measurement setup including 18 Arduino boards stacked up in two layers and power switch circuits.}
			\label{fig:Measure2}
		\end{subfigure}
		\caption{Measurement setup for the long-term test.}
		\label{fig:measurement}
		\vspace{-3mm}
	\end{figure}

	As shown in Fig. \ref{fig:measurement}, the whole measurement setup consists of the following components: 1) An external power supply that provides power to the setup system; 2) 2 Arduino boards as master boards. The two master boards are denoted as M0 and M1 respectively in Fig. \ref{fig:Measure1}. Each master board controls its slave boards via I2C protocol and communicates with other components in the system; 3) 16 Arduino as slave boards. The slave board reads out its SRAM power-up data and sends them back to its master board; 4) A power switch board that provides power to all the slave boards according to the command of each master board. Separate connections between the power switch and each slave board avoid interference between boards in the same stack; 5) Raspberry Pi receives SRAM data from master boards, and sends them to a database and stores them in a JSON format.
	
In Fig. \ref{fig:Measure1}, all the 18 Arduino boards were stacked up in two layers. The top layer consisting of M1 and from S16 to S23 is visible in Fig. \ref{fig:Measure2}. Two layers communicate with each other via connection between M0 and M1 so that data from different layers are synchronized, i.e. each slave board always produces the same quantity of SRAM PUF data within a fixed period of time. The layer at bottom is denoted as layer 0, and the layer at the top is denoted as layer 1. The detailed test flow of layer 0 as an example is described in Algorithm \ref{euclid}. 
\begin{figure}[tbp]
	\centerline{\includegraphics[width=0.4\textwidth]{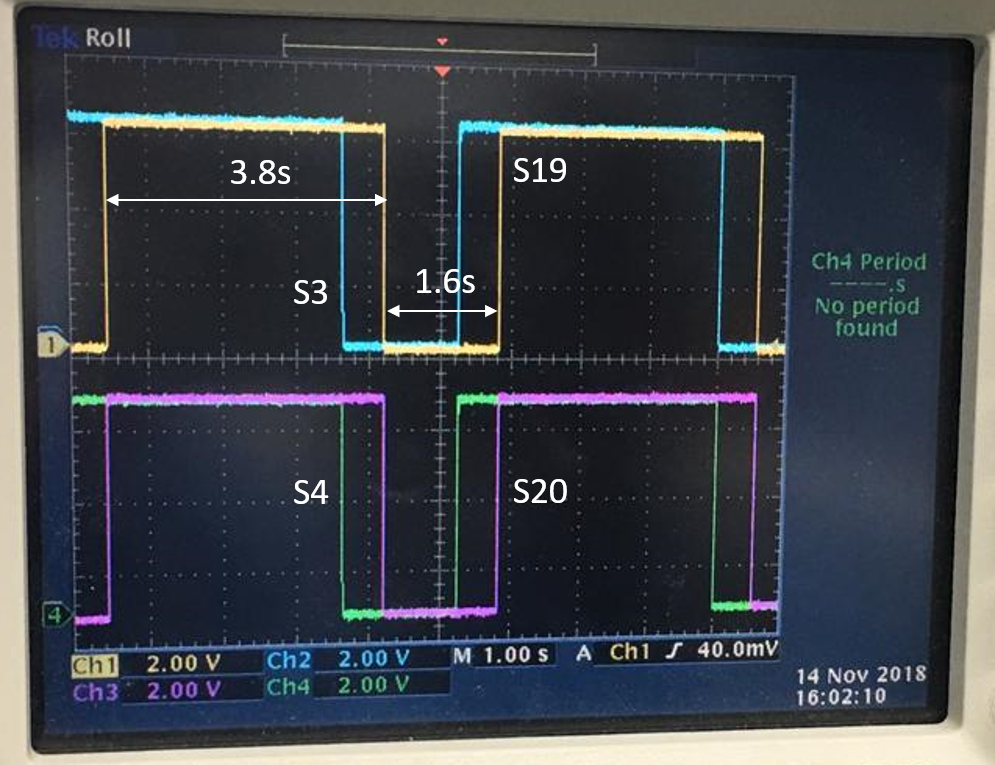}}
	\caption{Waveforms of power curves of board S3, S4, S19, S20 observed from the oscilloscope. }
	\label{fig:waveform}
\end{figure}

\begin{algorithm}[htbh]
	\begin{enumerate}
		\item M0 waits the end signal from M1, meaning layer 0 starts its test flow after layer 1 finalizes its last test cycle.
		\item M0 enables the power of its slave boards S0 to S7, via power switch
		\item When all the slaves on layer 0 are switched on, M0 gives the signal to M1, enabling layer 1 to starts its new test flow
		\item Each slave board reads the first 1 KByte of its SRAM power-up pattern, and sends the data back to M0 via I2C protocol
		\item M0 receives data from slave boards and sends the data to Raspberry pi
		\item M0 disables the power of slave boards S0 to S7
		\item M0 waits the start signal from M1, meaning layer 0 finalizes its cycle when autonomous read-out starts in layer 1.
		\item M0 gives the signal to M1, indicating that layer 0 finalizes its test cycle.
	\end{enumerate}
	\caption{Test flow on layer 0}\label{euclid}
\end{algorithm}
We used Tektronix TDS 3034B oscilloscope to observe the power cycle curve of four slave boards, namely S3, S4 from layer 0, and S19, S20 from layer 1. S3 and S19, S4 and S20 are in the same stack, respectively. As shown in Fig. \ref{fig:waveform}, the period of one power cycle is 5.4s. The power-on time of each board is 3.8s and the power-off time is 1.6s. Boards on the same layer (e.g. S19 and S20) are powered on and off at the same time. The power curves between two layers are unsynchronized to avoid interference, and to increase the
throughput of measurements.

\section {Evaluation}\label{evaluation}
\subsection{Initial SRAM PUF quality evaluation}
\begin{figure}[tbp]
	\centerline{\includegraphics[width=0.35\textwidth]{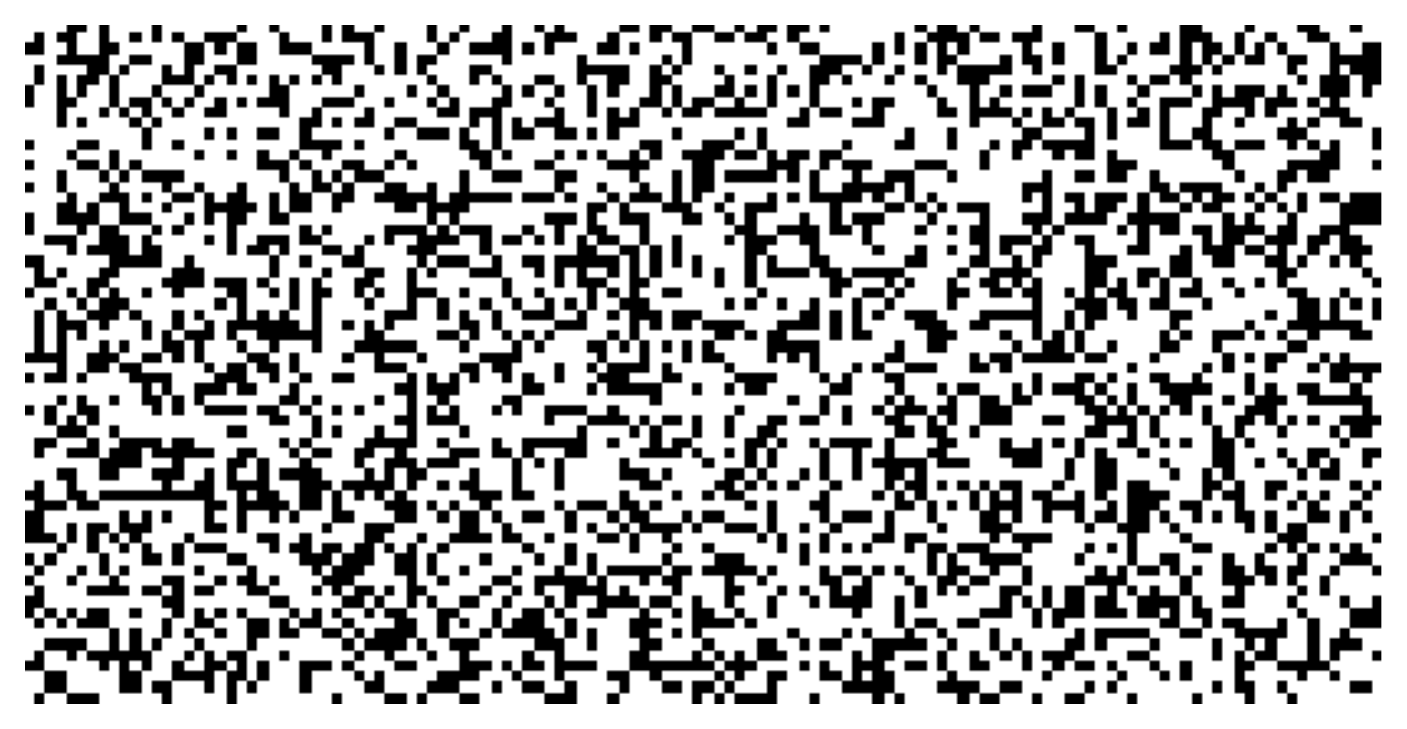}}
	\caption{Visualized startup pattern of 1KB memory on Arduino board with ID = 0.}
	\label{fig:Enrol}
	\vspace{-3mm}
\end{figure}

To visualize the distribution of the initial SRAM PUF, the first SRAM power-up pattern of board S0 is plotted in Fig. \ref{fig:Enrol}. For evaluating the initial SRAM PUF quality,  we select the first 1,000 read-out data of each board measured on Feb. 8, 2017 (starting date of the test). Three important metrics are adopted for the evaluation, as described in the following paragraphs.

\subsubsection {Within-Class Hamming Distance}
\begin{figure}[tbp]
	\centerline{\includegraphics[width=0.45\textwidth]{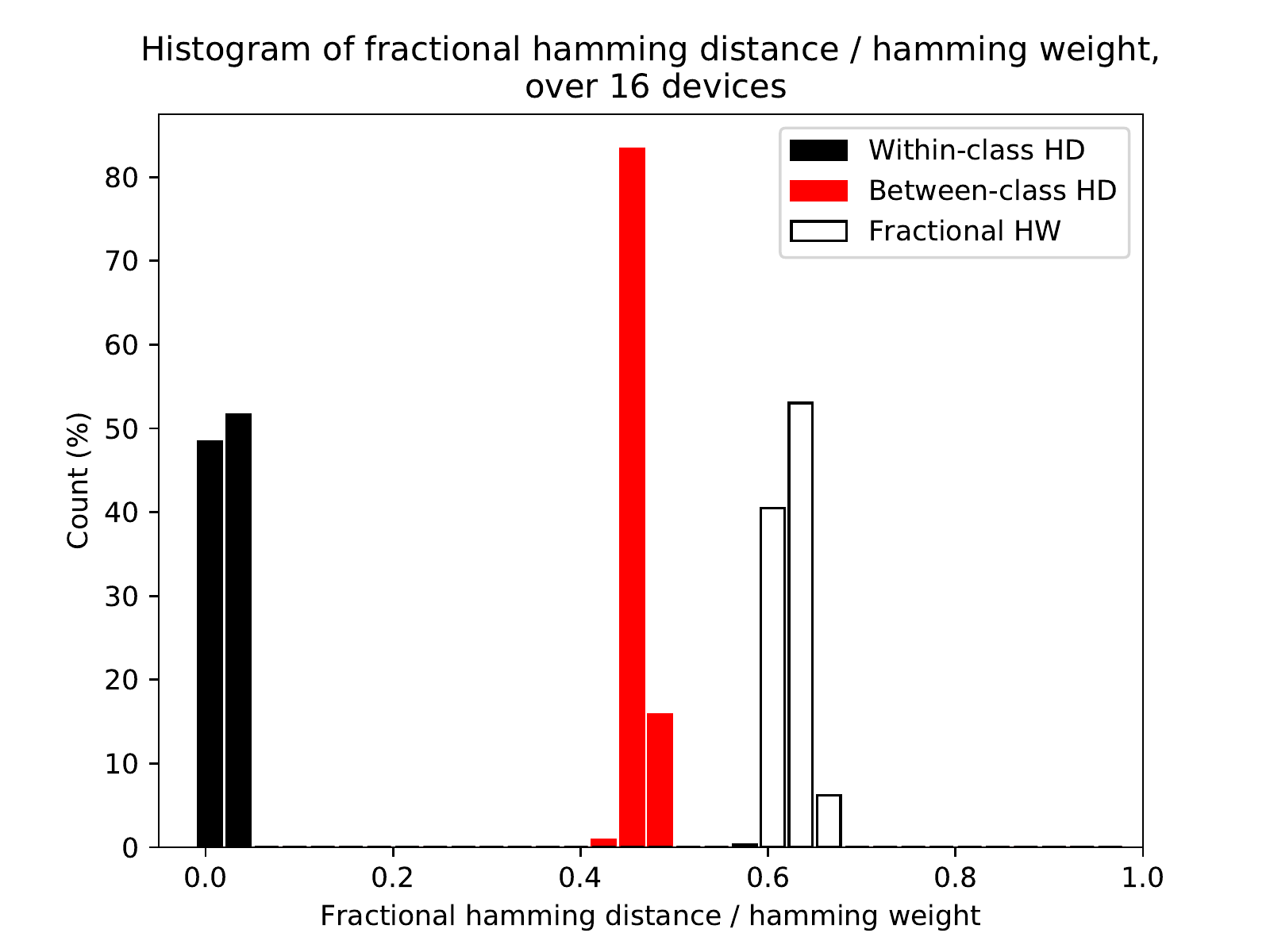}}
	\caption{Fractional Hamming distance / Hamming weight distributions at the beginning of the test (all the 16 Arduino boards included).}
	\label{fig:HD}
	\vspace{-3mm}
\end{figure}

 As described in Section \ref{SRAM PUF}, the difference between a reference pattern and another measured pattern of the same SRAM chip is supposed to be limited. To evaluate the reliability of SRAM PUF, the first read-out pattern is used as the reference. The other measurements of the same chip are compared to this reference using fractional Hamming Distance (FHD)\footnote{Hamming Distance (HD) is defined as the number of bits that different between two bit strings. In case of fractional Hamming Distance (FHD), the HD is divided by the length of the string.}. The FHD assessment within the same SRAM chip is defined as Within-Class Hamming Distance (WCHD). It is observed in Fig. \ref{fig:HD} that WCHD remains below 3\%. This value is well within the boundary, as specified in Section \ref{SRAM PUF} for error correction. So WCHD results show good reliability.
\subsubsection{Between-Class Hamming Distance} 
The SRAM PUF response of a chip should be unpredictable, even given the responses of other chips. The FHD between reference of every two different chips is defined as Between-Class Hamming Distance (BCHD). The BCHD is supposed to be significantly larger than WCHD and ideally close to 50\% so that the PUF of a certain chip is distinguishable from others. As shown in Fig. \ref{fig:HD}, the BCHD is distributed between 40\% and 50\%. The BCHD significantly deviates from the WCHD, showing good uniqueness among devices.

\subsubsection{Fractional Hamming Weight} 
Hamming Weight (HW) is the number of non-zero bits in the PUF response. In case of Fractional Hamming Weight (FHW), the HW is divided by the length of the string.  As shown in Fig. \ref{fig:HD}, the FHW of different SRAM chips lies between 60\% and 70\%. This indicates that the SRAM PUF is not perfectly unbiased, but still capable of generating a key of sufficient security strength \cite{Roel4}.

\subsection{Reliability and Uniqueness Evaluation}
We select the first 1,000 consecutive measurements after midnight on the 8th of each month for each SRAM chip. These SRAM PUF data are used for evaluating the effect of aging on reliability and uniqueness. The method for the evaluation is presented below.
\subsubsection{WCHD}We select the first read-out pattern of each board on the starting date as the reference. Measurements derived in the following months are compared to this reference.
\subsubsection{BCHD}For each month, the first SRAM read-out data of the 1,000 consecutive measurements mentioned above is used to calculate BCHD.
\subsubsection{FHW}For each month, we use the 1,000 consecutive measurements to calculate FHW.
\subsubsection{PUF entropy}In addition to the metrics mentioned above, PUF entropy is introduced to assess the uniqueness of SRAM PUFs using min-entropy \cite{entropy}. If a binary source has probability $p_0$ and $p_1$ to produce `0' and `1' respectively, then the definition for min-entropy of this binary source is:
\begin{equation*}
H_{min,PUF}=-\log_{2}{\left ( \max\left ( p_0,p_1 \right ) \right )}.
\end{equation*} 
Assuming all bits from the SRAM PUF are independent \cite{independency}, each bit location can be regarded as an individual binary source. Thus the average min-entropy of SRAM PUFs over $n$ bits is: 
\begin{equation*}
\left ( H_{min,PUF} \right )_{average}=\frac{1}{n}\sum_{i=1}^{n}-\log_2\left ( {\max\left ( p_{i\_0},p_{i\_1} \right )} \right ).
\end{equation*}
For the calculation, the SRAM PUF data for BCHD computation is used to derive PUF entropy, i.e. $p_0$ and $p_1$ are computed as probabilities over all measured SRAMs.
\begin{figure*}[ht]
	\vspace{-1ex}%
	\centering
	\begin{subfigure}[t]{0.35\textwidth}
		\centering
		\captionsetup{justification=centering,margin=0.0cm}
		\includegraphics[width=1.1\textwidth]{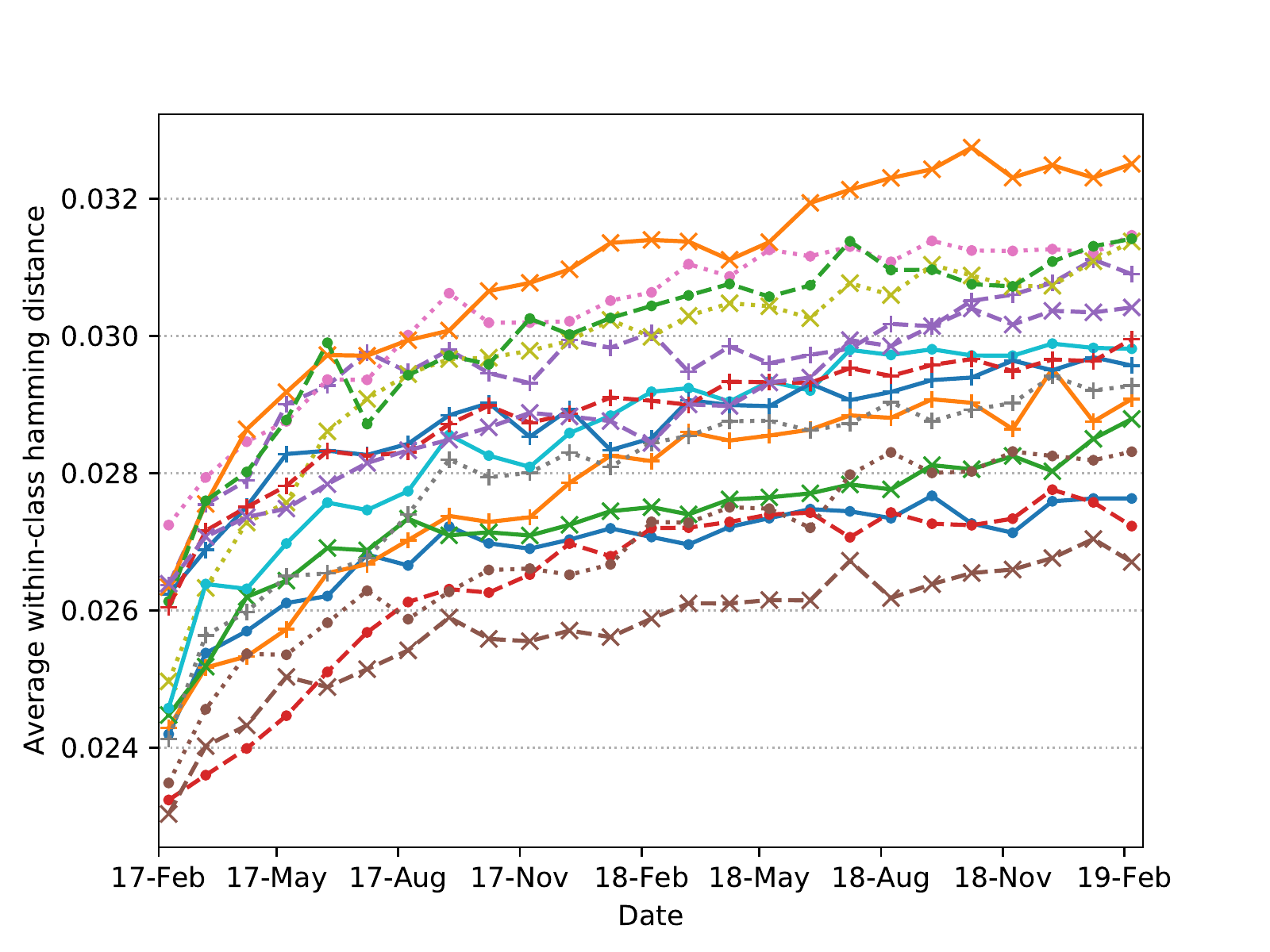}
		\caption{Development of WCHD with respect to the first SRAM observation at the start of the test}
		\label{fig:WCHD}
	\end{subfigure}
	\begin{subfigure}[t]{0.35\textwidth}
		\centering
		\captionsetup{justification=centering,margin=0.0cm}
		\includegraphics[width=1.1\textwidth]{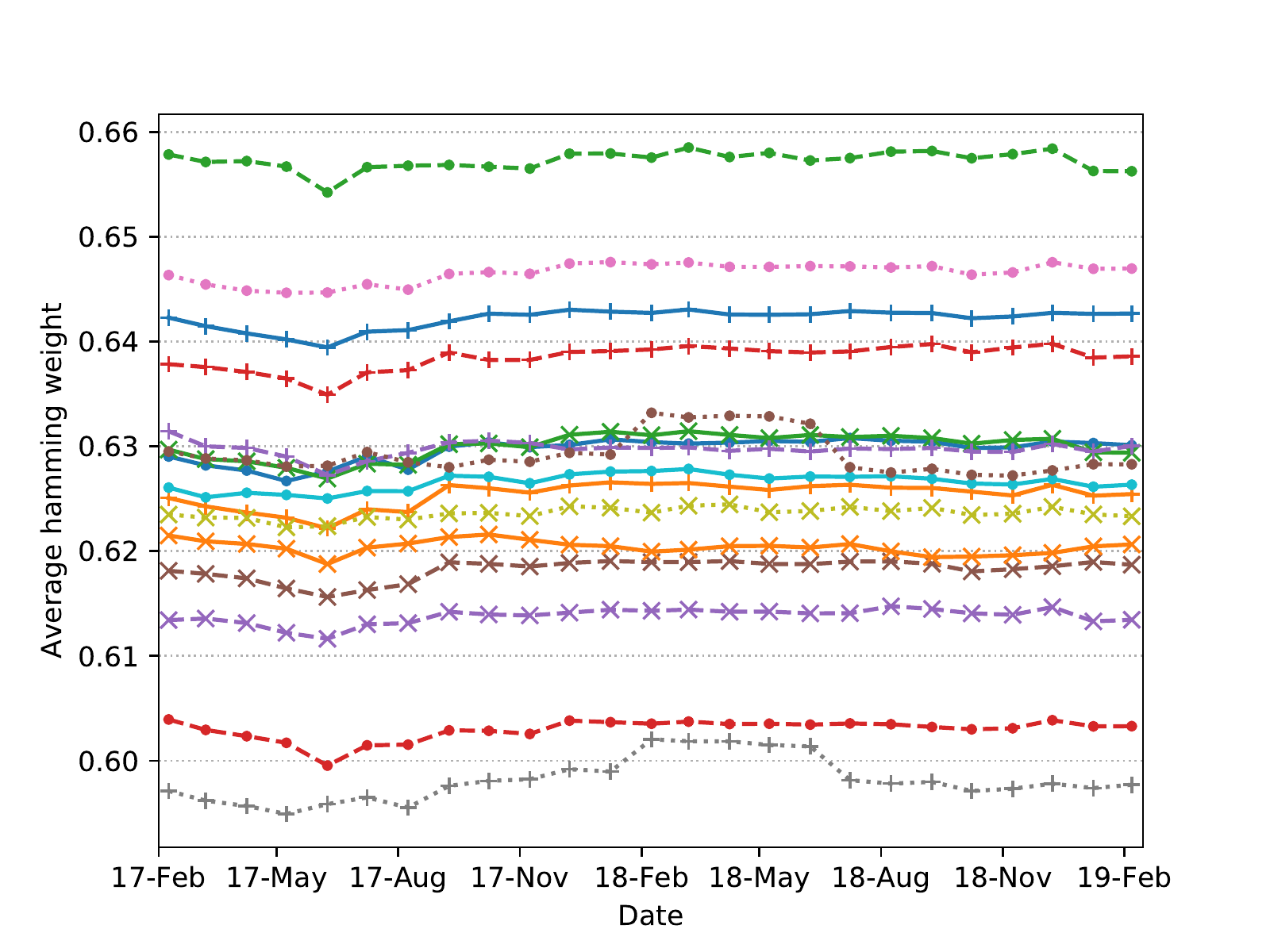}
		\caption{Development of Hamming weight}
		\label{fig:HW}
		
	\end{subfigure}	
	\begin{subfigure}{0.35\textwidth}
		\centering
		\captionsetup{justification=centering,margin=0.0cm}
		\includegraphics[width=1.1\textwidth]{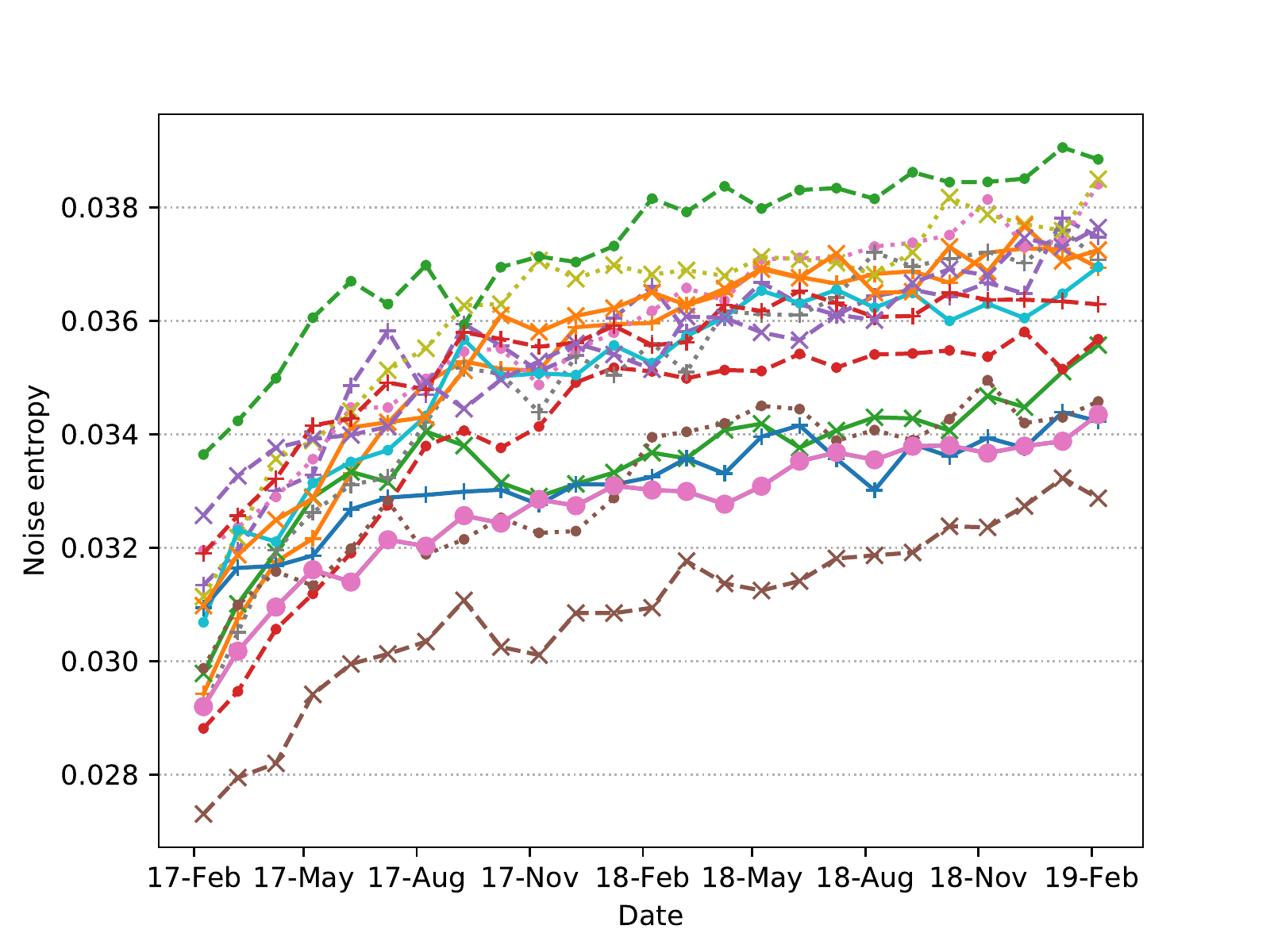}
		\caption{Development of noise entropy}
		\label{fig:random}
	\end{subfigure}
	\begin{subfigure}{0.35\textwidth}
		\centering
		\captionsetup{justification=centering,margin=0.0cm}
		\includegraphics[width=1.1\textwidth]{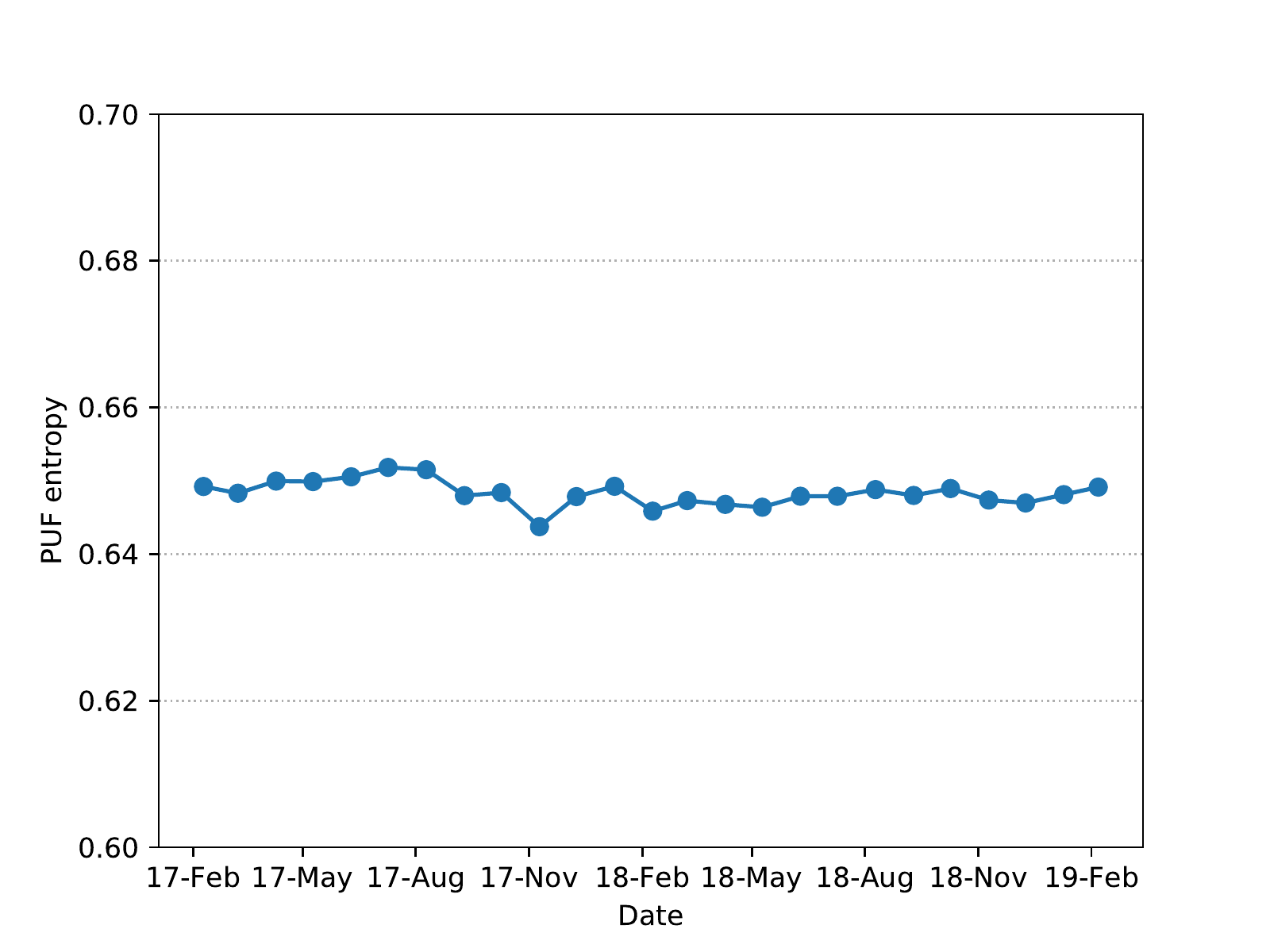}
		\caption{Development of PUF entropy}
		\label{fig:entropy}
	\end{subfigure}
	\caption{Development of reliability and uniqueness qualities of SRAM PUFs and randomness of SRAM PUF based TRNG on all the 16 Arduino boards during the long-term aging test. The aging period started from Feb. 2017 and ended in Feb. 2019. The test condition is at room temperature. Different lines in (a), (b) and (c) mean qualities of different SRAMs. }
	\label{fig:summary}
	\vspace{-3mm}
\end{figure*}
\subsection{Randomness Evaluation}
Apart from what are important if SRAM PUFs are applied for key generation, randomness is evaluated to assess the feasibility of using SRAM PUFs as a random source \cite{RNG}.  Two randomness-related properties are presented below.
\subsubsection{Stable cells}
Both process variation and electrical noise impact the skewness of SRAM cells \cite{Den}. An SRAM cell which is always powered up to state 0 or 1 over a large number of power-ups is considered as  a \textit{stable cell}. More quantitatively, one-probability is introduced to assess the stability of SRAM cells. One-probability ($p_i$) of a cell $i$ at a certain SRAM is the probability that the response value of this cell ($R_i$) is `1' over multiple power-ups \cite{Roel3}, defined as:
\begin{equation*}
p_i :=\textbf{Pr}(R_i=1)
\end{equation*}
In practice, the cell with one-probability of zero or one over 1,000 consecutive measurements in a certain month is counted as a stable cell in that particular month.
\subsubsection{Noise entropy} 
Noise entropy measures the unpredictability of next SRAM power-up response given the information of prior responses. A balanced cell is sensitive to noise and thus contributes to noise entropy of SRAM PUF based TRNG. Similar to PUF entropy, min-entropy is applied to estimate entropy of noise on SRAM PUF over multiple power-ups \cite{RNG}. Using 1,000 consecutive measurement data, the average min-entropy of noise on SRAM PUF over $n$ bits based on the one-probability of each bit is
\begin{equation*}
\left ( H_{min,noise} \right )_{average}=\frac{1}{n}\sum_{i=1}^{n}-\log_2\left ( {\max\left ( p_i,1-p_i \right )} \right ).
\end{equation*}

Note that $H_{min,noise}$ is noise entropy calculated based on multiple measurements of single SRAM, evaluating the randomness of TRNG based on SRAM PUFs. $H_{min,PUF}$ is PUF entropy calculated based single measurement of multiple SRAMs, evaluating the uniqueness of SRAM PUFs.
\subsection{Result discussion}\label{discussion}
The experimental aging results after 2-year of continuous test is summarized in Table \ref{tab1}. Both the average value (AVG.) and worst-case value (WC.) among 16 devices are presented in the table. The development of core properties over time is plotted in Fig. \ref{fig:summary}. The evaluation result of aging effect on SRAM PUF is discussed in two aspects.
\subsubsection{SRAM PUF as key generation scheme}
Within the time frame of 2-year aging, the FHD with respect to the initial reference at the start of the test is averagely increased from 2.49\% to 2.97\%, or in other words, the average number of bit errors with respect to the initial measurement increases by about 0.74\% each month over a 2-year period, indicated in Fig. \ref{fig:WCHD}. The increase of WCHD is in line with NBTI aging effect on SRAM. Therefore, the reliability of SRAM PUF worsens within a limited boundary due to aging effect. In contrast to accelerated aging showing WCHD increases from 5.3\% to 7.2\%, which is about 1.28\% monthly change over the first 2 years \cite{Roel2}, the reliability suffers less reduction. This means the previous research based on accelerated aging overestimated the degradation of SRAM PUF reliability.

Over the aging period, the HW of each SRAM remains almost constant, leading to a negligible change in BCHD and PUF entropy. Thus, the uniqueness of SRAM PUF is not impacted by aging effects. The results are indicated in Fig. \ref{fig:HW} and Fig. \ref{fig:entropy}.
\subsubsection{SRAM PUF as true random number generation}
The ratio of stable cells decreases from 85.9\% to 83.7\% on average among devices, with a 0.11\% decrease monthly. This result is also consistent with the hypothesis of NBTI aging effect in Section \ref{NBTI}.

As a consequence of fewer stable SRAM cells, the noise entropy is increased from 3.05\% to 3.64\%, with a 0.74\% increase rate per month. Therefore, the randomness of SRAM PUF based TRNG is improved after aging. The result is indicated in Fig. \ref{fig:random}.

Another notable result is that the monthly change rate in WCHD and noise entropy is larger at the start of the test than after 1 year. This can be explained as follows: We define stable cells as fully-skewed cells, and unstable cells with preference on power-up pattern as partially-skewed cells. For instance, in Fig. \ref{fig:SRAM}, assume that initially the SRAM cell is fully skewed at state zero, i.e. $V_{\mathrm{th,P2}}<V_{\mathrm{th,P1}}$ (all the values are treated as positive). At the start, the power-up state is always zero, leading to smaller $|V_{\mathrm{th,P2}}-V_{\mathrm{th,P1}}|$, which gradually converts this cell to a partially-skewed cell. This indicates that after a certain period of aging, the stable cell becomes unstable, and sometimes the new power-up state $Q=1$, leading to larger $|V_{\mathrm{th,P2}}-V_{\mathrm{th,P1}}|$. As a result,
 the tendency of $|V_{\mathrm{th,P2}}-V_{\mathrm{th,P1}}|$ is not monotonic over the aging. 
 	\begin{table}[tbp]
	\setlength{\tabcolsep}{+5pt}
	\caption{\textsc{Evaluation result of SRAM PUF qualities at the start and the end of the test}}
	\begin{center}
		{\renewcommand{\arraystretch}{1.35}
			\begin{tabular}{ l*6  c  }
				\hline
				\multicolumn{2}{c}{\textbf{Evaluation}} & \textbf{Start} & \textbf{End} &
				\begin{tabular}[c]{@{}c@{}}\textbf{Relative}\\ \textbf{Change}\end{tabular}&  
				\begin{tabular}[c]{@{}c@{}}\textbf{Monthly}\\ \textbf{Change}\end{tabular}\\
				\hline
				\multirow{2}{*}{\textbf{WCHD}} & AVG. & 2.49\% & 2.97\% & +19.3\% & +0.74\%\\
				&  WC. & 2.72\% & 3.25\% & +19.5\% & +0.74\%\\
				\multirow{2}{*}{\textbf{HW}} & AVG. & 62.70\% & 62.70\% & negligible$^{\mathrm{a}}$ & negligible  \\
				&  WC. & 65.78\% & 65.62\% & -0.24\% & -0.01\%\\
				\multirow{2}{*}{\begin{tabular}{@{}l@{}}\textbf{Ratio of}\\\textbf{Stable Cells}\end{tabular}} & AVG. & 85.9\% & 83.7\% & -2.49\% & -0.11\%\\
				&  WC. & 87.2\% & 85.4\% & -2.22\% & -0.87\%\\
				\multirow{2}{*}{\textbf{Noise entropy}} & AVG. & 3.05\% & 3.64\% & +19.3\% & +0.74\%\\ 
				&  WC. & 2.73\% & 3.29\% & +20.5\% & +0.78\% \\
				\multirow{2}{*}{\textbf{BCHD}} & AVG. & 46.79\% & 46.80\% & negligible & negligible\\
				& WC. & 44.31\% & 44.67\% & +0.81\% & 0.03\%\\
				\textbf{PUF entropy} & & 64.92\% & 64.91\% & negligible & negligible\\
				\hline
				\multicolumn{6}{l}{$^{\mathrm{a}}$Negligible means change is less than 0.01\%.}
			\end{tabular}
		}
		\label{tab1}
	\end{center}
\vspace{-4mm}
\end{table}
\section{Conclusion}\label{conclusion}
This paper has presented the result of silicon aging on SRAM as PUF and as source of TRNG based on long-term test in nominal conditions.
 Based on the experimental result derived from 2-year measurement, the evaluation confirms the anticipated NBTI effect on SRAM cells. The uniqueness keeps almost constant during the aging period. The reliability of SRAM PUF gradually degrades, the extent of which, is less
 pessimistic than the emulated result from accelerated aging test. WCHD increases by 0.74\% each month in nominal conditions, while WCHD increases by 1.28\% each month in accelerated aging.  Meanwhile, the aging effect slightly improves the randomness of SRAM PUF based TRNG. Noise entropy improved by 0.74\% each month. 

\section*{acknowledgment}
The author would like to thank Peter Simons and Tobias Adryan for their help in establishing and maintaining the measurement setup. This work has been supported by H2020 Project ``INSTET\footnote{https://cordis.europa.eu/project/rcn/216842/factsheet/en}: Securing the Internet of Things with a unique microchip fingerprinting technology" with grant agreement No 811509.

\end{document}